\documentclass[aps,twocolumn,english,superscriptaddress,citeautoscript,preprintnumbers,amsmath,amssymb,floatfix,footinbib]{revtex4-1}
\usepackage{graphicx}
\usepackage{graphics}               
\usepackage{bm}   
\usepackage{amssymb } 
\usepackage{array}       
\usepackage[breaklinks=true,colorlinks,citecolor=blue,linkcolor=blue,urlcolor=blue]{hyperref}
\usepackage{float}
\usepackage[table]{xcolor}
\definecolor{Cyan}{rgb}{0.4,1,0.7}
\definecolor{LightCyan}{rgb}{0.8,1,0.85}

\begin{document}
\title{Experimental and theoretical study of the correlated compound YbCdSn: Evidence for large magnetoresistance and mass enhancement}
\author{Antu Laha}
\affiliation{Department of Physics, Indian Institute of Technology, Kanpur 208016, India}
\author{P. Rambabu}
\affiliation{Department of Physics, Indian Institute of Technology Hyderabad, Kandi, Medak 502 285,
	Telengana, India}
\affiliation{Department of Pure and Applied Physics, Guru Ghasidas Vishwavidyalaya, Koni, Bilaspur 495009, India}
\author{V. Kanchana}
\affiliation{Department of Physics, Indian Institute of Technology Hyderabad, Kandi, Medak 502 285,
	Telengana, India}
\author{L. Petit}
\affiliation{Science and Technology Facilities Council, Daresbury Laboratory, Daresbury WA4 4AD, United Kingdom}
\author{Z. Szotek}
\affiliation{Science and Technology Facilities Council, Daresbury Laboratory, Daresbury WA4 4AD, United Kingdom}	
\author{Z. Hossain}
\email{zakir@iitk.ac.in}
\affiliation{Department of Physics, Indian Institute of Technology, Kanpur 208016, India}

\begin{abstract}
The unusual features of topological semimetals arise from its nontrivial band structure. The impact of strong electron correlations on the topological states remains largely unexplored in real materials. Here, we report the magnetotransport properties of YbCdSn single crystals. We found two fundamental experimental evidences of electron correlations through magnetic susceptibility and specific heat. The electron correlations in this compound lead to an intermediate valence state and enhance the effective mass of the charge carriers. This correlated state exhibits large nonsaturating magnetoresistance, low carrier density, magnetic field induced metal-semiconductor-like crossover and a plateau in resistivity at low temperatures. This compound also shows a cusp-like magnetoconductivity at low magnetic field which indicates the presence of weak antilocalization effect. Our band structure calculations of Yb$^{2+}$ state predict YbCdSn to be a topological nodal-line semimetal.
\end{abstract}
\maketitle
\section{Introduction}
Strongly correlated electron systems and topological semimetals are two separate active frontiers of modern condensed matter research but their coexistence is sparse. The low-energy electronic excitations in topological semimetals are characterized by Dirac-like linear band dispersion, yielding a variety of exotic transport phenomena such as extremely large nonsaturating magnetoresistance (MR), chiral anomaly induced negative longitudinal MR and anomalous Hall effect \cite{RevModPhys_WSMs,Review_WSMs,RevModPhys_TIs}. On the other side, strong electron correlations give rise to several other remarkable features like valence fluctuations, Kondo effect and heavy fermion behaviour \cite{Heavy_fermion,Kondo_effect,Valence_fluctuation,Ce2Ni3Ge5_JMMM_2018}.  

Since the topological states are robust against perturbations and local disorder, the appearance of electron correlations makes this states more intriguing. For example, the electron correlations in the perovskite oxide CaIrO$_3$ tune the Dirac line node to the proximity of Fermi level which yields highly mobile electrons \cite{CaIrO3_natcomn_2019,CaIrO3_PRL_2019}. Moreover, correlation induced unconventional mass enhancement of charge carriers was observed at very high magnetic field around the Dirac nodal loop in the topological semimetal ZrSiS \cite{ZrSiS_NaturePhysics_2018}. 

The interplay between topology and electron correlations can be better visualized in rare-earth based topological materials where couplings between the $f$-electrons and conduction electrons are prominent. In these systems, strong correlation renormalizes the nontrivial electronic bands which exhibit strong $f$-character and enhance the effective mass of the charge carriers \cite{TKI_PRL_2010}. The $d$-$f$ hybridization opens a bulk band gap at low temperature due to electron correlations which lead to a topological Kondo insulator state in SmB$_6$ \cite{SmB6_PRL_2014,SmB6_PRX_2013}. Moreover, theory predicted few correlated topological phases with gapless excitations in rare-earth based compounds such as Weyl heavy fermion state and Weyl-Kondo semimetal phase \cite{PNAS_YbPtBi,CeRu4Sn6_PRX_2017}. Among them, experimental evidence was found in YbPtBi, where Weyl fermion state is pronounced at high temperatures while $f$-electrons are well localized, whereas strong Kondo interaction renormalizes the bands and leads to a heavy fermion state at low temperatures \cite{YbPtBi_NatureCom_2018}.

In the present work, we report the presence of electron correlations in YbCdSn through magnetic susceptibility and specific heat measurements. This compound hosts an intermediate valence state in which significant mass enhancement of the charge carriers is observed. The magnetotransport studies reveal several novel transport properties such as large nonsaturating magnetoresistance, low carrier density, magnetic field induced metal-semiconductor-like crossover and resistivity plateau at low temperatures. Weak antilocalization effect is also observed at low magnetic fields and low temperatures. The main objective of the electronic structure calculations, based on a first-principles methodology, is to establish the energy ground state and electronic configuration (valence) of the system and with that inspect details of the corresponding density of states (DOS) and band structure, in the vicinity of the Fermi energy, in search of a possible topological nodal-line semimetal state in YbCdSn.

\section{Methods}
\subsection{Experimental details}
Single crystals of YbCdSn were synthesized using cadmium flux \cite{YbCdGe_PRB_2019,CaCdSn_PRB_2020}. Yb ingot (99.99$\%$, Alfa Aesar), 
Cd shots (99.99$\%$, Alfa Aesar) and Sn pieces (99.99$\%$, Alfa Aesar) in molar ratio of 1:47:1 were mixed in an alumina crucible, 
then the crucible was sealed into a quartz tube under partial pressure of argon gas. The content was heated to 1000$^\circ$C, 
kept for 6 hours at that temperature, and then cooled to 500$^\circ$C at a rate of 3$^\circ$C/hour \cite{YbCdGe_PRB_2019,CaCdSn_PRB_2020}. 
Needle-like single crystals were extracted from the flux by centrifuging. The crystals are air sensitive in nature and they require handling in inert 
atmosphere. The crystal structure and the phase purity were determined by x-ray diffraction (XRD) technique using Cu-K$_\alpha$ 
radiation in a PANalytical X$'$Pert PRO diffractometer. Magnetotransport measurements were carried out in a physical property 
measurement system (PPMS, Quantum Design) via standard four-probe method. Magnetic susceptibility and specific heat were also 
measured in the same PPMS using the vibrating sample magnetometer and relaxation method respectively. 

\subsection{Computational details}
Two different approaches, both based on density functional theory (DFT)~\cite{hohenberg1964,kohn1965}, are combined here to study the electronic structure of YbCdSn. The first one, used mainly for exploring the correlated nature of the YbCdSn ground state and the corresponding valence state of Yb ion, is the self-interaction corrected local spin density approximation (SIC-LSDA) to exchange-correlation energy functional (see APPENDIX for details)~\cite{PZ1981,Temmerman_Handbook,Petit_REX}, implemented in the linear muffin-tin orbitals method with the atomic sphere approximation (LMTO-ASA)~\cite{OKA1975}, where the polyhedral Wigner Seitz cell is approximated by slightly overlapping atom centered spheres, with a total volume equal to the actual crystal volume. The second approach makes use of the generalized gradient approximation (GGA) to the exchange-correlation energy functional, with an empirical effective Hubbard U parameter (GGA+U), added to account for the Yb $f$-electron correlations, as implemented in the Vienna {\it ab-initio} simulation package (VASP). It is employed mostly for optimizing the YbCdSn crystal structure and its band structure calculation.

The first-principles band structure calculations are performed using the projector augmented wave (PAW) method from 
VASP package~\cite{T1,T3,T4,T5,T6}. The generalized gradient approximation of Perdew, Burke and Ernzerhof (GGA-PBE) \cite{T7} is used to describe the exchange and correlation effects in the system. The strong correlation effects of rare earth Yb-$f$ states are treated by applying an effective Hubbard U value of 6.0 eV to Yb-$f$ states, using  Dudarev's approach \cite{T8}. The cutoff value of plane wave kinetic energy is set as 516 eV and a k-mesh of  $8\times8\times12$ grid is chosen using Monkhorst-Pack scheme \cite{T9}, while the Brillouin zone integration is done using Gaussian broadening \cite{T10} of 0.05 eV in the calculations. The equilibrium of the system is achieved with a force convergence of 10$^{-3}$ eV/\AA $~$and the self-consistent energy criteria is set as 10$^{-6}$ eV/cell. For the surface state calculations, firstly tight-binding Hamiltonian is obtained based on maximally localized Wannier functions using Wannier90 package \cite{Pizzi}. The orbital projections chosen for Yb, Cd and Sn atoms respectively are $s,p,d,f$; $s,p$; and $s,p$. Based on this tight-binding model, the iterative Green's function method \cite{wu2018wanniertools, sancho1985highly} which is implemented in WannierTools \cite{wu2018wanniertools} package is used to investigate the topological properties such as surface state spectrum, Fermi arc and Z2 indices.

\section{Crystal structure}
\begin{figure}
	\centering
	\includegraphics[width=0.9\linewidth]{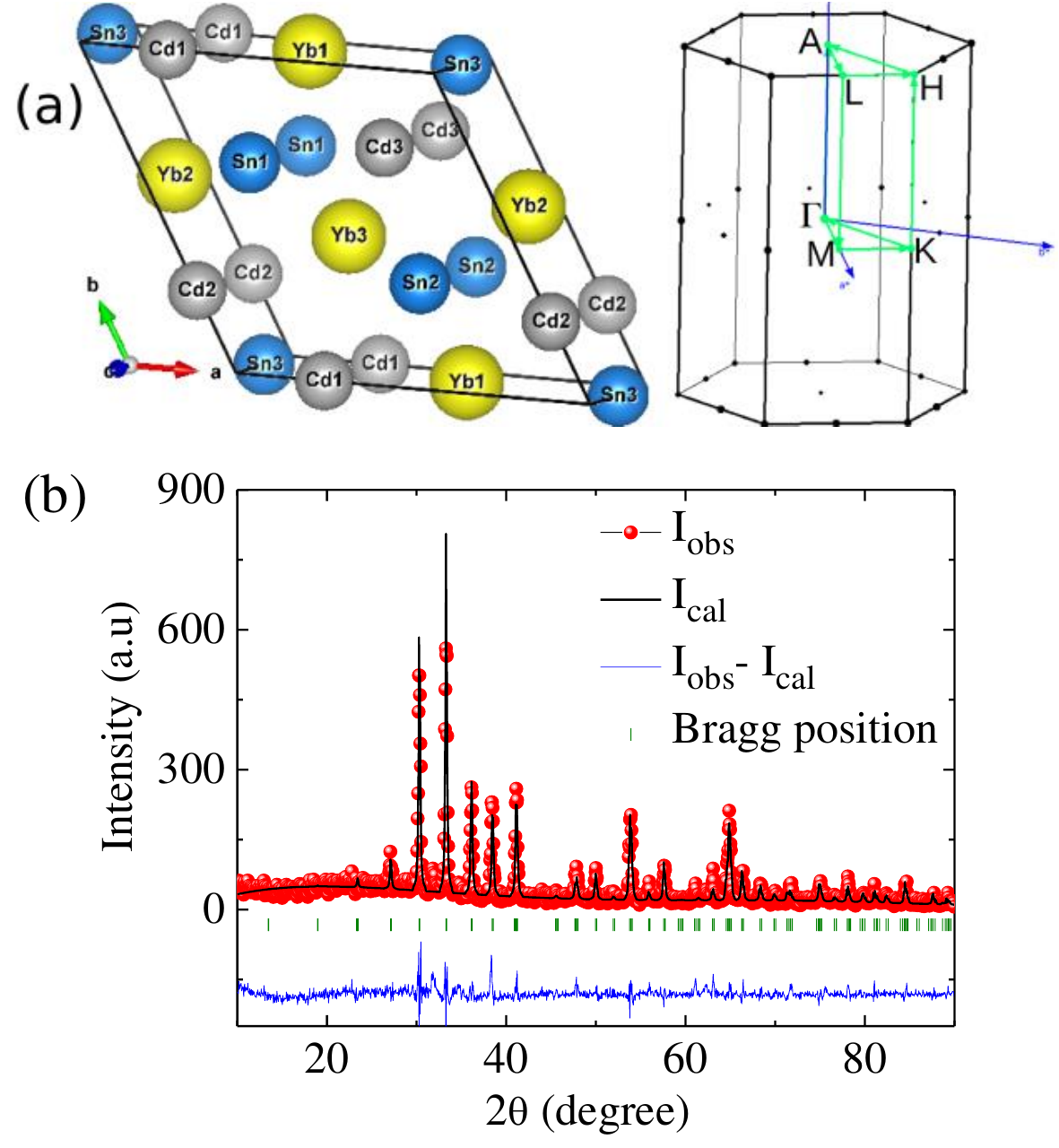}
	\caption{(a) Crystal structure and Brillouin zone of YbCdSn. (b) Powder x-ray diffraction pattern of crushed YbCdSn single crystals, recorded at room temperature. The observed intensity (red scattered points), Rietveld refinement fit (solid black line), difference between the experimentally observed and calculated intensities (solid blue line) and Bragg peak positions (vertical green bars) are shown. 
		\label{XRD}}
\end{figure}

YbCdSn hosts a similar hexagonal crystal structure (space group $P$-$62m$) like the family of nodal-line semimetals RCdX (R=Ca,Yb; X=Ge,Sn) \cite{CaCdGe_PRB_2017,CaCdSn_PRB_2020,YbCdGe_PRB_2019}. The crystal structure and the Brillouin zone are shown in Fig.\ref{XRD}(a), where three Yb, three Cd, two Sn1, and one Sn2 atoms are located at 3$g$ ($x$, 0, 1/2), 3$f$ ($x$, 0, 0), 2$c$ (1/3, 2/3, 0), and l$b$ (0, 0, 1/2) positions respectively. The Rietveld structural refinement of the powder XRD data was done using the FULLPROF software package \cite{Fullprof_1993} as shown in Fig.\ref{XRD}(b). No impurity phase was observed within our experimental resolution. The Rietveld refinement yields a hexagonal structure with space group  $P$-$62 m$ (No. 189) and the refined lattice parameters are $a$ = $b$ = 7.5937(7) \AA $~$and $c$ = 4.6790(2) \AA.

\section{Experimental results}
\subsection{Magnetic susceptibility}
\begin{figure}
	\centering
	\includegraphics[width=0.99\linewidth]{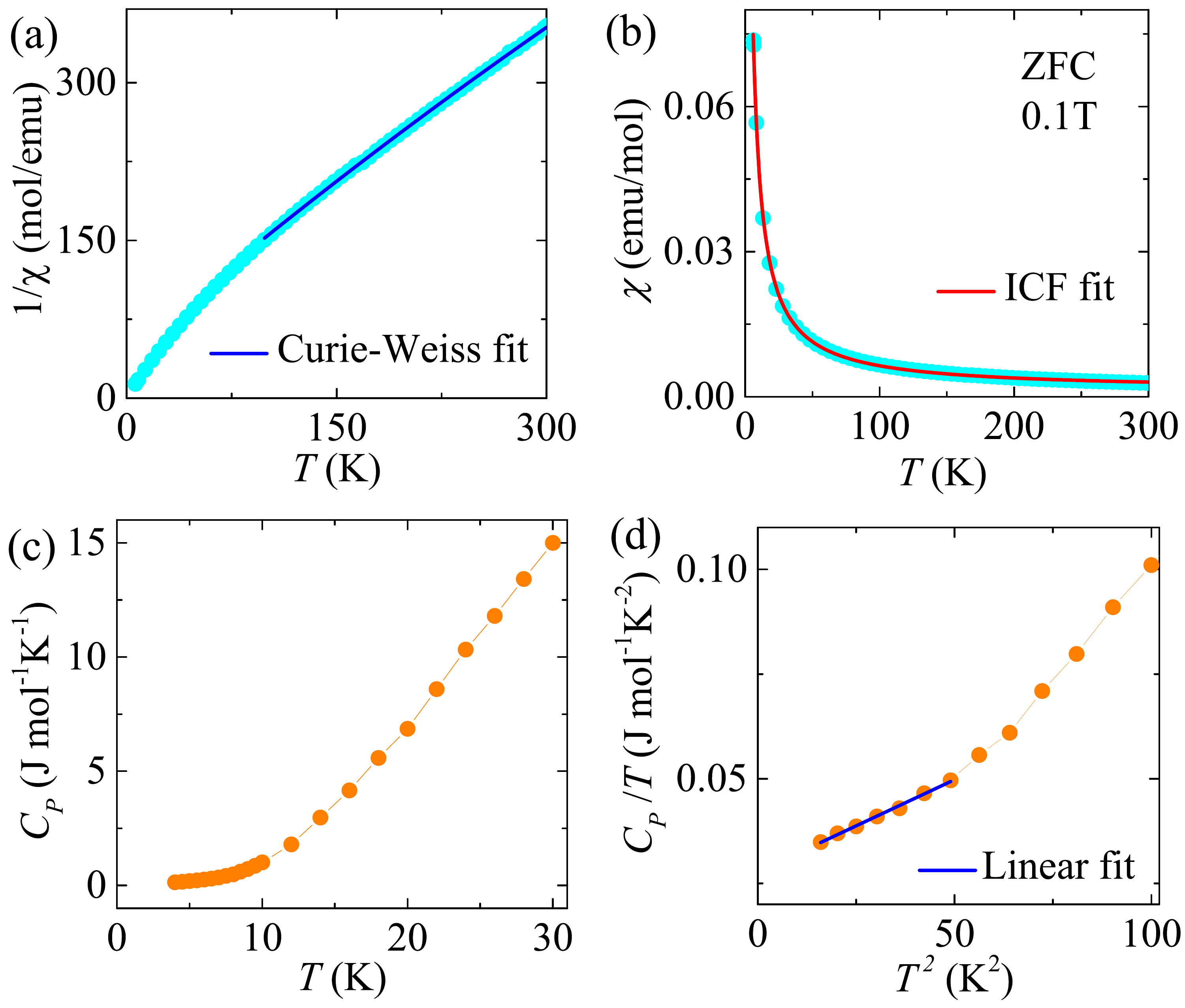} 
	\caption{(a) The Curie-Weiss law fitting (solid blue line) to the inverse magnetic susceptibility (zero field cooled) in the temperature range 100 K-300 K ($B\parallel c$). (b) The solid red line fits to the two-level ionic interconfiguration fluctuations model (Eq.(\ref{equ1})). (c) Temperature dependence of the specific heat from 4 K to 30 K. (d) Linear fitting to the curve $C_P/T$ vs. $T^2$ in the temperature range 4 K-7 K.
		\label{fig1}}
\end{figure}

To find out the magnetic ground state of Yb in YbCdSn, we measured temperature dependent zero field cooled (ZFC) magnetic 
susceptibility ($\chi$) of the grown single crystal along the crystallographic $c$-axis ($B \parallel c$). The 
inverse magnetic susceptibility $1/\chi$ fits to the modified Curie-Weiss law, $\chi (T) = \chi_0 + C/(T - \theta)$ in the 
temperature range 100 K-300 K where $C$, $\theta$, $\chi_0$ are Curie constant, paramagnetic Curie temperature and the temperature 
independent part of magnetic susceptibility respectively [Fig.\ref{fig1}(a)]. The effective magnetic moment of Yb ions 
estimated from Curie constant is $\mu_{eff}=2.61 \mu_B$ which lies in between the values for Yb$^{2+}$ state ($0 \mu_B$) and Yb$^{3+}$ 
state ($4.54 \mu_B$) and the nature of inverse magnetic susceptibility curve is typical for intermediate valence compounds \cite{VF_YbAlB4,YbCdGe_PRB_2019}. 

For more insight, we use two-level ionic interconfiguration fluctuations model (ICF) to describe the magnetic susceptibility data. According to ICF model, the overall $\chi(T)$ is given by \cite{Hirst1970,Sales_and_Wohlleben,Franz}
\begin{equation} \label{equ1}
\chi(T) = \frac{N}{3k_B}\Big[\frac{\mu_n^2~ \nu(T) + \mu_{n-1}^2 (1- \nu(T))} {T^*}\Big]
 +f \frac{C}{T-\theta}+\chi_0
\end{equation}
with
\begin{equation}
\nu(T) = \frac{2J_n + 1}{(2J_n + 1)+(2J_{n-1}+1)~\textrm{exp}(-E_{ex}/k_B T^*)}
\end{equation}  
where $T^* = (T + T_{sf})$. Here, $\nu(T)$ is the fractional occupation of ground state, $\mu_n$ and $\mu_{n-1}$ are the effective moments in $4f^n$(Yb$^{2+}$) and $4f^{n-1}$(Yb$^{3+}$) states, $(2J_n + 1)$ and $(2J_{n-1}+1)$ are the degeneracies of the corresponding energy states of $E_n$ and $E_{n-1}$, $E_{ex}$ = ($E_n$ - $E_{n-1}$) is the interconfigurational excitation energy, $T_{sf}$ is the spin fluctuation temperature associated with the valence fluctuation and $f$ is the fraction of the contribution originating from stable Yb$^{3+}$ ions. In the equation \ref{equ1}, first term represents valence fluctuation part, second term is for stable Yb$^{3+}$ state and $\chi_0$ is the temperature independent part. The fitting of equation \ref{equ1} gives a good agreement with the experimental data (red solid line in Fig.\ref{fig1}(b)). The fitting parameters are comparable with several other Yb-based intermediate valence compounds \cite{YbCuGa,YbNiAl4,Kaczorowski2016}. Therefore, the result indicates that YbCdSn exhibits a strongly correlated electron phenomenon which is valence fluctuation.

\subsection{Specific heat}
To get further evidence for electron correlation, we measured temperature dependent specific heat ($C_P$) as shown in Fig.\ref{fig1}(c). The specific heat follows the expression $C_P=\gamma T+\beta T^3$ in the temperature range 4 K-7 K. The estimated value of $\gamma$ from the linear fit to the curve $C_P/T$ vs. $T^2$ is 28 mJ mol$^{-1}$ K$^{-2}$ [ Fig.\ref{fig1}(d)]. This value is larger than that observed in non-correlated systems but it is comparable to that observed in intermediate valence compounds such as YbAl$_3$ ($\gamma$ = 58 mJ mol$^{-1}$ K$^{-2}$), YbFe$_2$Al$_{10}$ ($\gamma$ = 35 mJ mol$^{-1}$ K$^{-2}$) and YbFe$_4$Sb$_{12}$ ($\gamma$ = 140 mJ mol$^{-1}$ K$^{-2}$) \cite{YbAl3_PhysicaB_2000,YbFe2Al10_PRB_2017,YbFe4Sb12_PRB_1998}. The large $\gamma$-value in these compounds indicates the enhancement of effective mass due to the interaction between $4f$ and conduction electrons. 
\begin{figure*}
	\centering
	\includegraphics[width=0.90\linewidth]{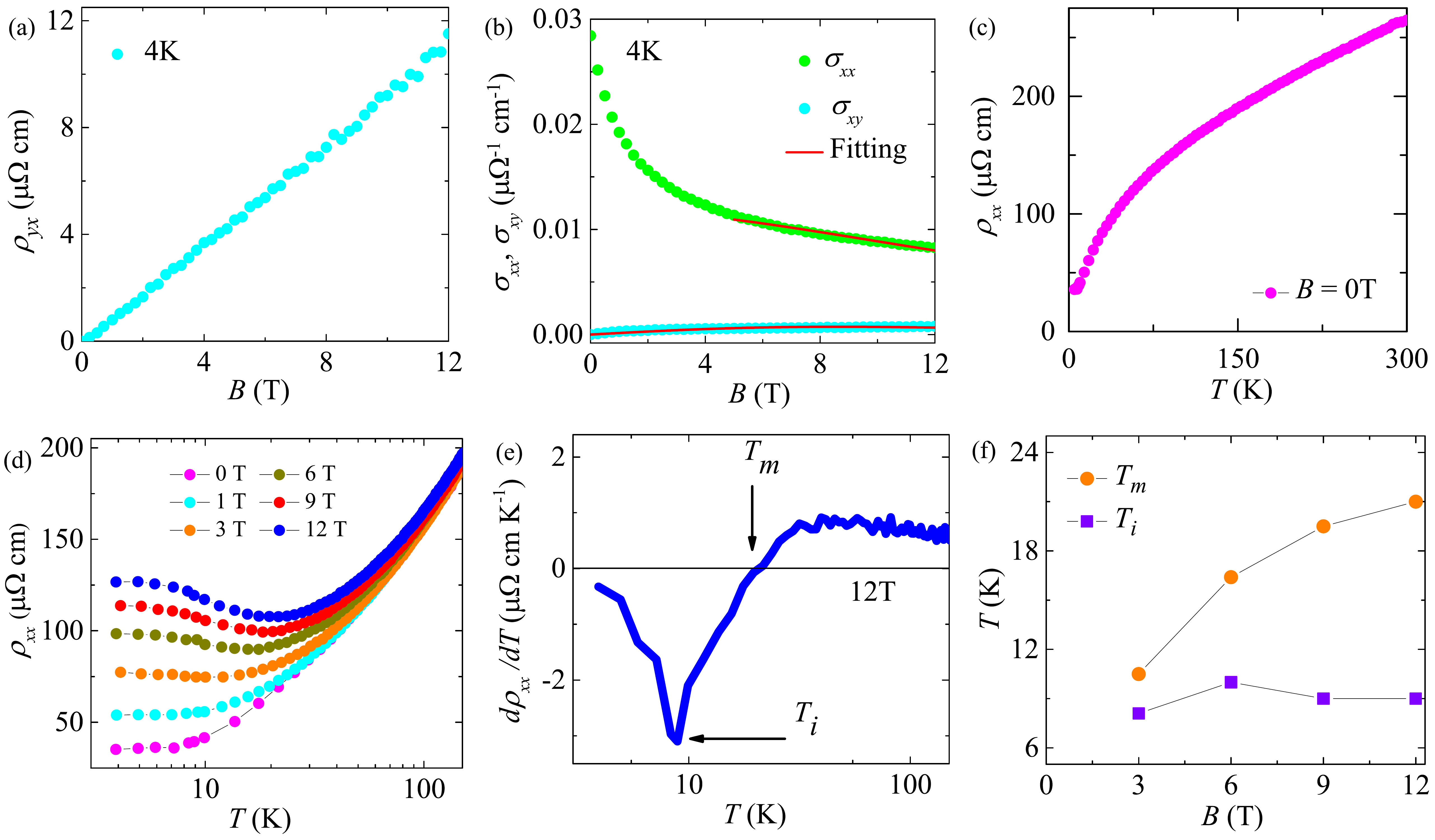} 
	\caption{(a) Hall resistivity ($\rho_{yx}$) as a function of $B$. (b) Global fitting of Hall conductivity ($\sigma_{xy}$) and longitudinal conductivity ($\sigma_{xx}$) at 4 K using two band model (equation \ref{xx} and \ref{xy}). (c) Temperature dependence of electrical resistivity ($\rho_{xx}$). (d) Electrical resistivity as a function of $T$ for various $B$ up to 12 T.  (e) First derivative of resistivity $d\rho_{xx}/dT$ as a function of $T$. (f) The two characteristic temperatures $T_i$ and $T_m$ as a function of $B$. ($B\perp I$ and $I\parallel c$).
		\label{fig2}}
\end{figure*}
\subsection{Hall resistivity and longitudinal resistivity}
To determine the mobility and the density of the charge carriers in YbCdSn, we measured magnetic field dependent Hall resistivity ($\rho_{yx}$) at 4 K. The Hall resistivity is presented in Fig.\ref{fig2}(a) after removing the MR contribution using the expression $\rho_{yx}=[\rho_{yx}(B)-\rho_{yx}(-B)]/2$. The $\rho_{yx}$ shows linear field dependence behavior up to a high magnetic field of 12 T. In order to probe the electron-hole compensation in this compound, we consider the semiclassical two band model \cite{Two_Band_original},
\begin{equation}
\sigma_{xx}= e\Big[\frac{n_h \mu_h}{(1+\mu_e B)^2} + \frac{n_e \mu_e}{(1+\mu_h B)^2}\Big],
\label{xx}
\end{equation}
\begin{equation}
\sigma_{xy}= eB\Big[\frac{n_h \mu_h^2}{(1+\mu_h B)^2} - \frac{n_e \mu_e^2}{(1+\mu_e B)^2}\Big],
\label{xy}
\end{equation}
where $n_h$ ($n_e$) and $\mu_h$ ($\mu_e$) are the hole (electron) density and mobility respectively. The longitudinal conductivity and the Hall conductivity are obtained using the expressions $\sigma_{xx}=\rho_{xx}/(\rho_{xx}^2+\rho_{yx}^2)$ and $\sigma_{xy}=\rho_{yx}/(\rho_{xx}^2+\rho_{yx}^2)$ respectively. Simultaneous fitting of the $\sigma_{xx}$ and $\sigma_{xy}$ [Fig.\ref{fig2}(b)] yields $n_h= 7.08 \times 10^{19}$ cm$^{-3}$, $n_e= 5.96 \times 10^{19}$ cm$^{-3}$ and $\mu_h = 4.90 \times 10^2 $ cm$^2$V$^{-1}$s$^{-1}$, $\mu_e = 6.63 \times 10^2 $ cm$^2$V$^{-1}$s$^{-1}$. The $\sigma_{xx}$ deviates from the semiclassical two band model at low magnetic field due to the influence of weak antilocalization effect which is discussed later. The ratio $n_e/n_h$ is about 0.84 which indicates nearly perfect electron-hole compensation \cite{LuPtBi_PRB_2015}. Our theoretically calculated hole and electron densities ($n_h= 2.45 \times 10^{19}$ cm$^{-3}$, $n_e= 1.61 \times 10^{19}$ cm$^{-3}$) are comparable with the experimental values and also support the electron-hole compensation.

Now, we present the temperature dependence of electrical resistivity ($\rho_{xx}$) of YbCdSn under various magnetic fields as shown in Fig.\ref{fig2}(d). The $\rho_{xx}$ decreases with decreasing temperature down to 7 K and then saturates with a residual resistivity of $35.2~\mu\Omega$ cm [Fig.\ref{fig2}(c)]. Above a certain magnetic field ($\geq 3$ T), the $\rho_{xx}$ shows a field induced metal-semiconductor-like crossover and a resistivity plateau at low temperatures, which become more prominent at higher magnetic field. These features were observed in its sister compound CaCdSn and several other topological semimetals \cite{CaCdSn_PRB_2020,YbCdGe_PRB_2019,PhysRevLett.94.166601, LaSb_NaturePhy_2015,NbAs2TaAs2_PRB(R)_2016,TaSb2_PRB(R)_2016,LaSb_PRL_2016,PtBi2_PRL_2017,ZrSiS_ScienceAdv_2016,du2018}. To analyze these features, we plot first derivative of resistivity ($d\rho_{xx}/dT$) as a function of temperature as shown in Fig.\ref{fig2}(e). From this figure, two characteristic temperatures can be clearly identified - (i) the temperature associated with metal-semiconductor-like crossover ($T_m$) where $d\rho_{xx}/dT=0$, (ii) the inflection point in resistivity ($T_i$) where $d\rho_{xx}/dT$ is minimum. The resistivity plateau starts to appear just below $T_i$. Magnetic field dependent variations of $T_m$ and $T_i$ are shown in Fig.\ref{fig2}(f).

\subsection{Magnetoresistance}
\begin{figure}
	\centering
	\includegraphics[width=0.99\linewidth]{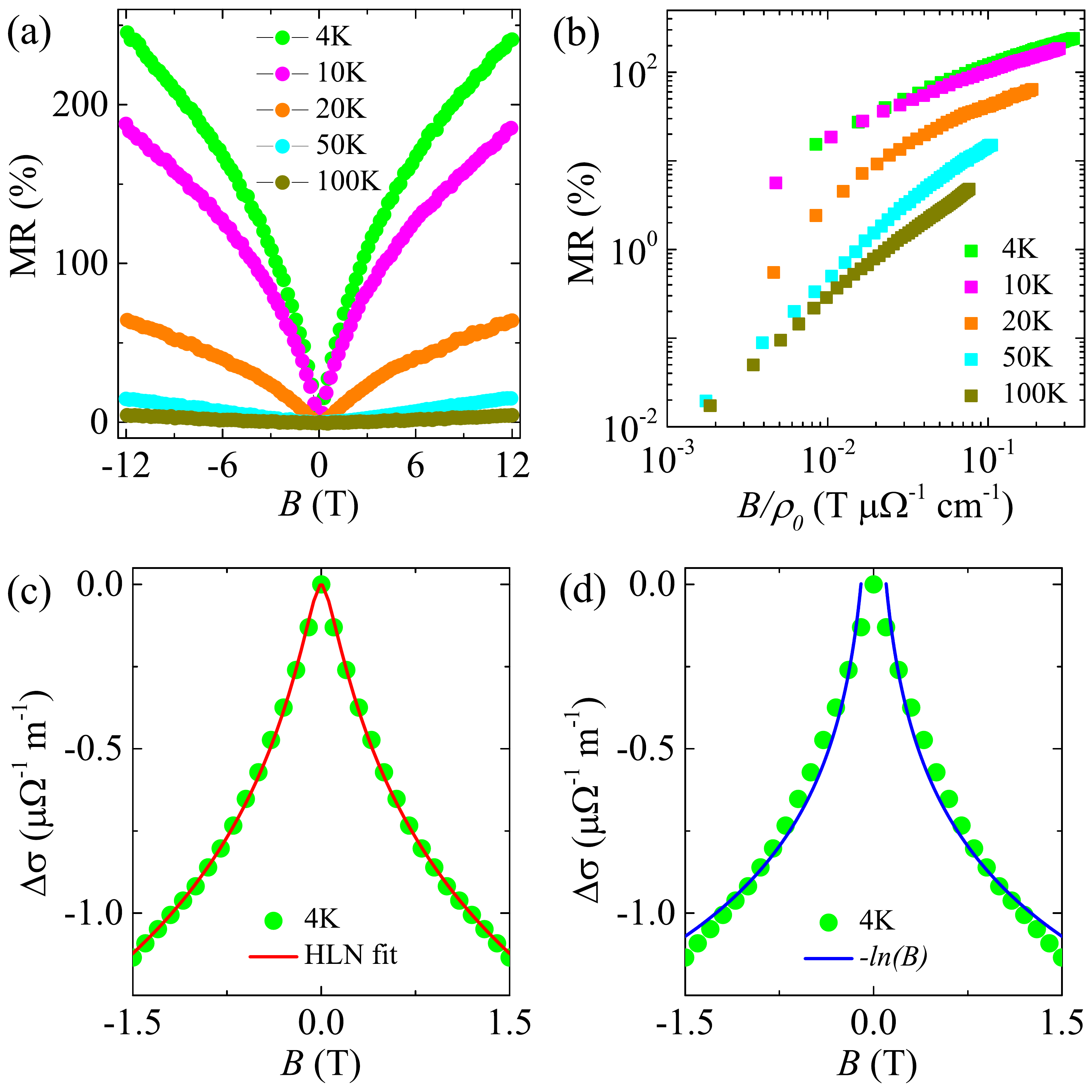} 
	\caption{(a) Magnetic field dependence of magnetoresistance [MR=$(\rho_{xx}(B)-\rho_{xx}(0))/\rho_{xx}(0)$] for various $T$ ($B\perp I$ and $I\parallel c$). (b) MR as a function of $B/\rho_0 \rightarrow $ Kohler's plot. (c) Magnetoconductivity [$\Delta \sigma=\sigma(B)-\sigma(0)$] as a function of $B$. The red solid line fits to the Eq.\ref{(equ3)} [HLN formula]. (d) Solid blue line shows $-ln(B)$ fit.}
	\label{fig3}
\end{figure}

The magnetoresistance of YbCdSn as a function of $B$ is shown in Fig.\ref{fig3}(a) for various $T$. A large non-saturating MR of $2.46\times 10^2\%$ is observed at 4 K and 12 T. The MR decreases significantly as we increase the temperature and reaches to a lower value of 4.4$\%$ at 100 K and 12 T but still no sign of saturation is observed. The possible origin of large MR in this compound is electron-hole compensation effect \cite{WTe2_PRB_2015,NbAs_PRB_2015,CaCdGe_PRB_2017}. To describe the MR, we also use Kohler's rule that is MR = $\alpha (B/\rho_0)^m$, where $\alpha$ is a constant. In Fig.\ref{fig3}(b), the MR data are plotted as a function of $(B/\rho_0)$ but the MR data at various temperatures are not merging into a single curve which indicates a clear violation of Kohler's scaling. Violation of Kohler's scaling can be observed due to various reasons including the existence of multiple carriers and different scattering rates of various carriers \cite{McKenzie_Kohler,WP_PRB_2017, Xu_2015,Multiband_Kohler}.

Furthermore, a cusp-like feature is observed in the MR at low magnetic field which is the reminiscent of weak antilocalization (WAL) effect. This effect was generally observed in 2D-materials but recently it was also found in 3D topological materials \cite{Bi2Te3_PRB_2017,LuPdBi_SR_2014,LuPtSb_AIP_2015,YbCdGe_PRB_2019}. To understand the WAL effect in our system, we analyzed the magnetoconductivity (MC) data using two different available models. One is well known Hikami-Larkin-Nagaoka (HLN) model \cite{HLN_2D}. According to this model, the magnetoconductivity for WAL effect can be expressed as
\begin{equation}\label{(equ3)}
\Delta \sigma = \frac{\alpha e^2}{2\pi^2\hbar}\Bigg[\Psi \Bigg(\frac{1}{2} + \frac{\hbar}{4el_\phi^2 B}\Bigg) - ln\Bigg(\frac{\hbar}{4el_\phi^2 B}\Bigg)\Bigg],
\end{equation} 
where $\Psi$ is the digamma function and $l_\phi$ is the phase coherence length. The parameter $\alpha$ gives an estimation for number of conduction channel participating in transport and $\alpha$ is found to be -1/2 per conduction channel in 2D electron systems \cite{Bi2Se3_PRL_2010,Bi2Te3_PRL_2011}. We calculate the MC [$\Delta\sigma=\sigma(B)-\sigma(0)$] using the expression $\sigma=\rho_{xx}/(\rho_{xx}^2+\rho_{yx}^2)$, where $\sigma(B)$ and  $\sigma(0)$ are the conductivity at finite magnetic field and at zero magnetic field respectively. The MC data fits well to the HLN formula in the field range -1.5 T $\leq B \leq$ 1.5 T as shown in Fig.\ref{fig3}(c). The fitting yields $\alpha\sim -10^4$ and $l_\phi \sim 67$ nm. Such a large value of $\alpha$ was also found in other 3D materials due to the presence of large number of conduction channels in the bulk \cite{Bi2Te3_PRB_2017,LuPdBi_SR_2014,LuPtSb_AIP_2015,YbCdGe_PRB_2019}.

Recently another model was proposed to describe the WAL effect in nodal-line semimetals. According to this model, the MC at low field varies as $-ln(B)$ for a torus-shaped Fermi surface \cite{WAL_PRL_2019}. A bad fitting of our MC data to this model suggests that our system does not belong to the above mentioned particular case [see Fig.\ref{fig3}(d)]. 

\section{Theoretical results}
\subsection{Ground state configuration and density of states}
\begin{figure}
        \centering
        \includegraphics[width=0.99\linewidth]{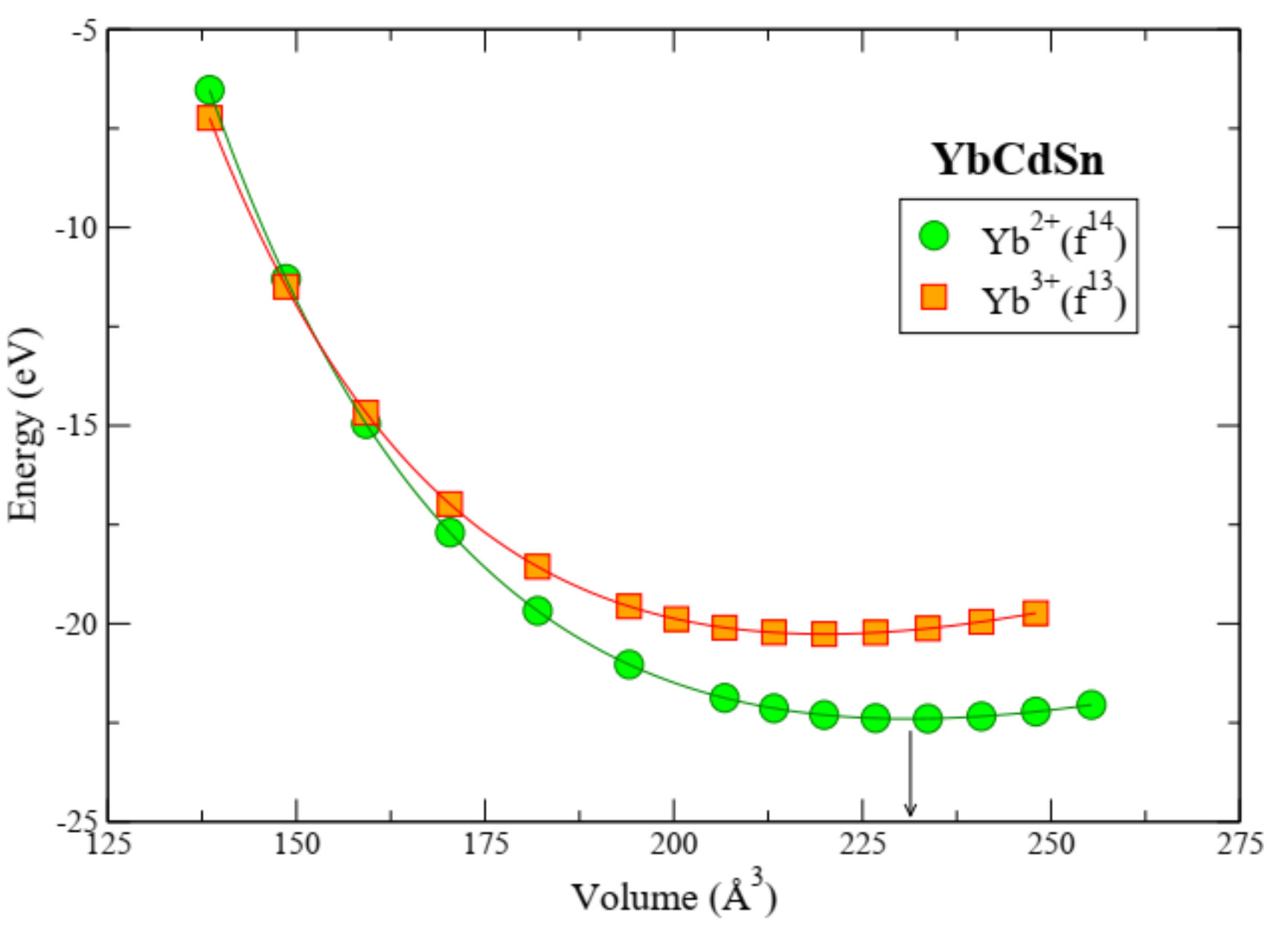}
        \caption{ Total energy as a function of volume for YbCdSn with respectively Yb$^{2+}$ and Yb$^{3+}$ configurations. The energy minima for the respective curves define their theoretical volumes.}
        \label{etot}
\end{figure}
\begin{figure}
        \centering
        \includegraphics[width=0.99\linewidth]{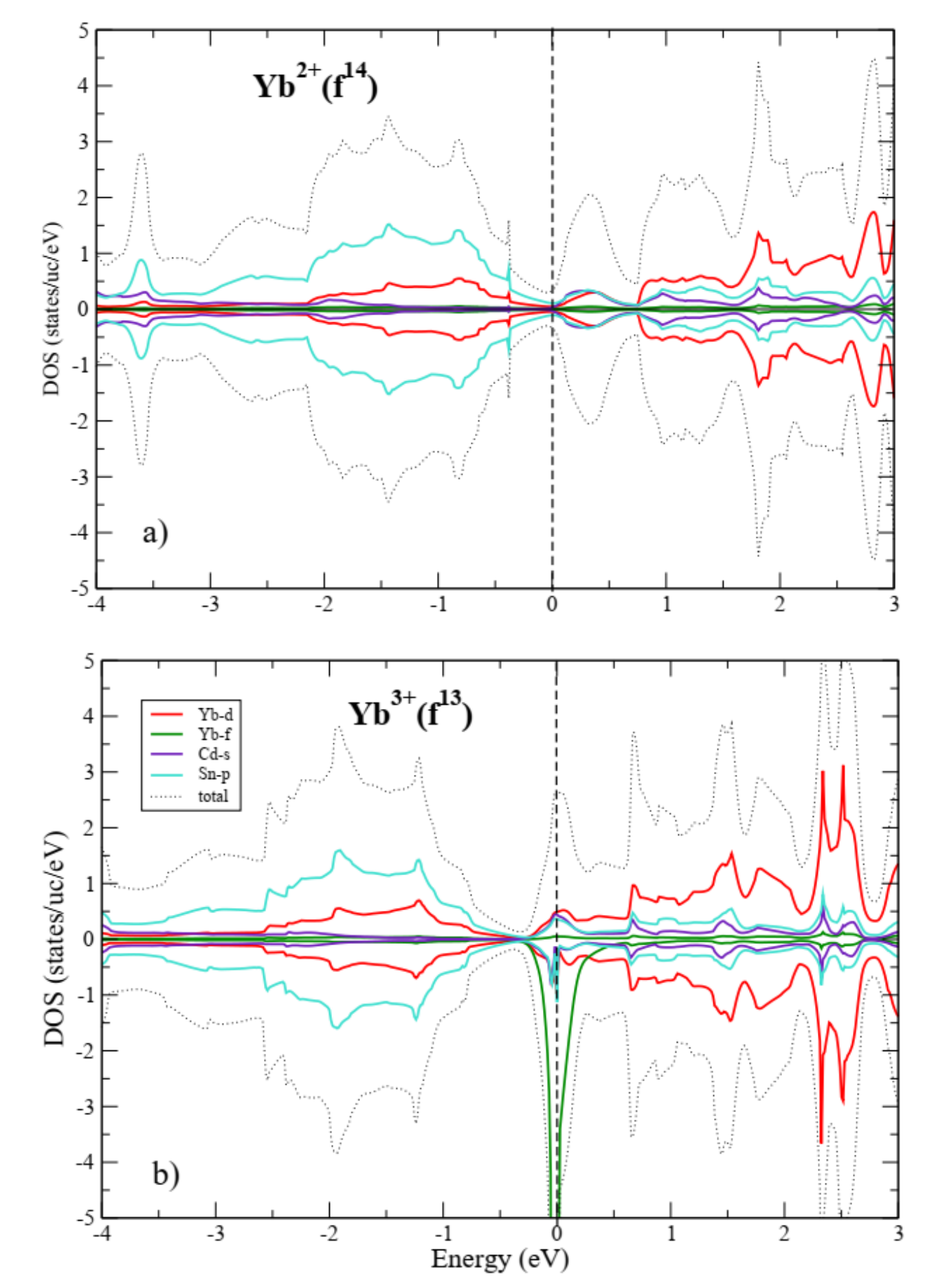} 
        
        \caption{The spin decomposed total densities of states of YbCdSn with respectively a) Yb$^{2+}$ and b) Yb$^{3+}$ configurations. Displayed are also partial densities of states for some of the electron states of Yb, Cd and Sn. The spin-up DOS is plotted on the positive side of the y-axis, and the spin-down DOS on its negative side. Energy equal to 0 eV marks the Fermi energy}
        \label{dos}
\end{figure}

In the lanthanide compounds the ground-state valence depends on the lanthanide ion as well as its chemical environment~\cite{Petit_REX}. Lanthanide ions are usually found in trivalent state, although in some compounds Yb-ions can be found in divalent and intermediate/mixed valence states~\cite{Strange_Nature1999,WMT1999,Svane_YbX_PRB}. Here for YbCdSn, we realize two relevant valence configurations, namely Yb$^{2+}$ with all 14 $f$-electrons treated as localized, and Yb$^{3+}$ with 13 localized $f$-electrons and one $f$-electron band-like, allowed to hybridize with the other valence electrons. For these two cases we performed the self-consistent SIC-LSDA total energy calculations as a function of volume and calculated also the corresponding electronic structure in terms of densities of states. The results of these SIC-LSD calculations are shown in Figs. \ref{etot} and \ref{dos}. As seen in Fig. \ref{etot}, the global energy minimum occurs in the fully localized, Yb$^{2+}$, scenario, at a volume of 231.2 \AA$^3$, only slightly below the experimental value. The Yb$^{3+}$ configuration is energetically less favourable by about 2 eV and has a spin magnetic moment of -0.65 $\mu_B$, while the Yb$^{2+}$ scenario has zero spin magnetic moment. The total energy versus volume behaviour of the LSDA configuration, with all $f$-electrons treated as itinerant, has also been calculated, but turns out to be energetically unfavourable by about 55 eV and is therefore not shown in Fig. \ref{etot}. Under pressure, with decreasing volume, a localization/delocalization transition from Yb$^{2+}$ to Yb$^{3+}$ is seen to occur at a rather considerable pressure of 54 GPa. Although this transition would seem rather difficult to realize in practice, the valence fluctuations between the Yb$^{2+}$ and Yb$^{3+}$ configurations could not be excluded, in line with the proposed intermediate-valent state of this system.

The spin-decomposed total densities of states of YbCdSn for both the Yb$^{2+}$ and Yb$^{3+}$ configurations are plotted respectively in Figs. \ref{dos}a and \ref{dos}b, in the vicinity of the Fermi energy. For both configurations, respectively 14 and 13 localized $f$ band states are not shown in the figure, and only a single band-like $f$ state for the Yb$^{3+}$ configuration is seen to sit right at the Fermi energy in the negative, spin-down panel of the figure. For the Yb$^{2+}$ configuration, the total density of states per spin is equal to 0.34 states/unit cell/eV, in agreement with a semimetallic character, while for the Yb$^{3+}$ configuration it is equal to 2.67 states/unit cell/eV for the spin-up component and 99.87 states/unit cell/eV for the spin-down component.

One should mention here that because the SIC-LSDA approach is still a one-electron ground state theory, with the screening/relaxation effects neglected, it does not give accurate removal energies of the localized $f$ states which, as a result, are usually situated at too low an energy (here around -10 eV and therefore are not shown in Figs. \ref{dos}a and \ref{dos}b). Of course, the position of the delocalized $f$-peak in the DOS for the Yb$^{3+}$ configuration is also affected by the neglect of the screening/relaxation effects which, if included, would shift it slightly above the Fermi energy (according to a $\Delta$SCF-like estimate~\cite{WMT_Pr}). Thus the $f$ character would still be present at the Fermi energy, through hybridization with the other conduction electron states, but its contribution to the spin-down density of states at the Fermi energy would be considerably reduced, making the spin-down DOS closer to the value of the spin-up DOS. Depending on the actual position of the delocalized $f$-peak relative to the Fermi level, the DOS of Yb$^{3+}$ configuration could be interpreted as either metallic or heavy Fermion-like.

\subsection{Electronic structure}
\begin{figure}
	\centering
	\includegraphics[width=0.99\linewidth]{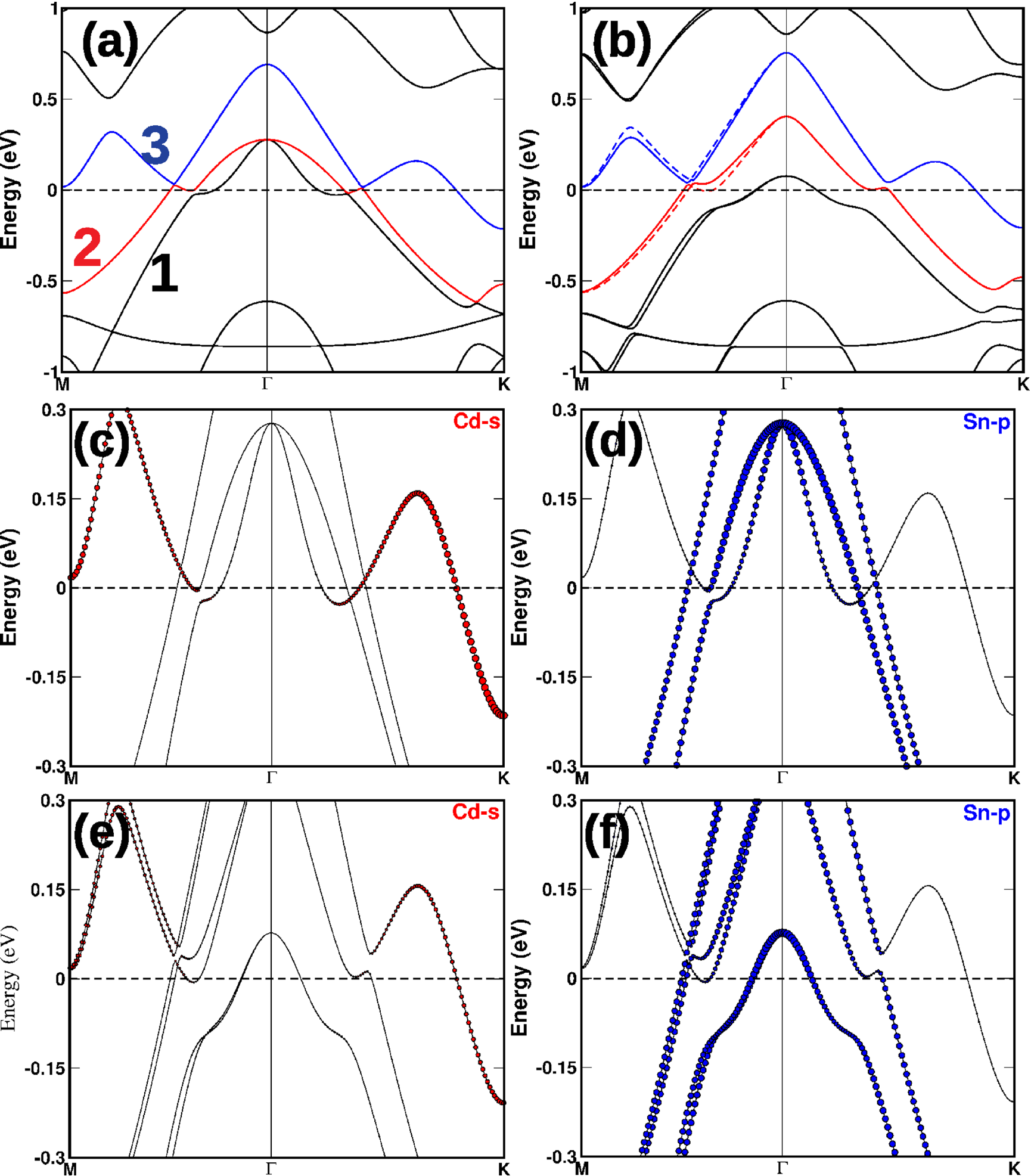} 
	\caption{Band structure of YbCdSn along $M-\Gamma-K$ path without SOC in (a), with SOC in (b). The character bands $Cd-s$ and $Sn-p$ are given without SOC in (c,d) and with SOC in (e,f) respectively. The number of bands that cross E$_F$ are marked as 1, 2 and 3 for without SOC case in (a).}
	\label{fig4}
\end{figure}
\begin{figure}
	\centering
	\includegraphics[width=0.9\linewidth]{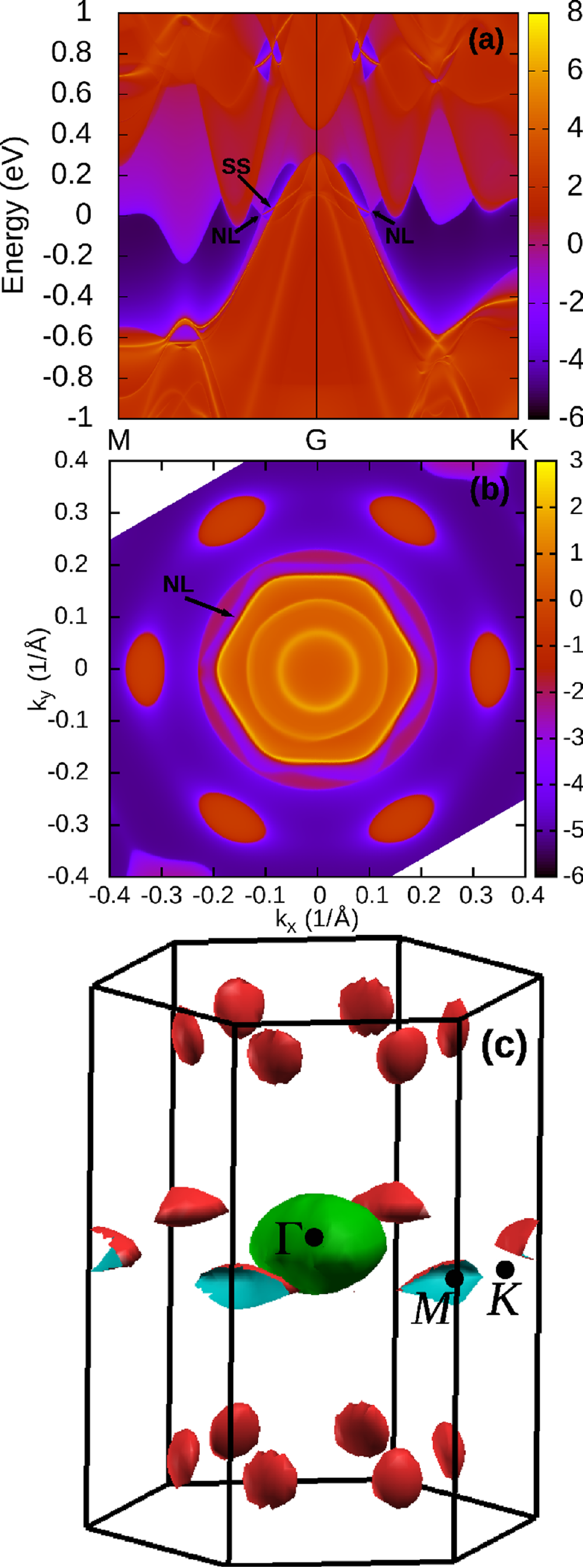} 
	\caption{(a) The (001) projected surface states of YbCdSn without SOC and (b) the corresponding Fermi arc in $k_z$=0 plane. (c) The Fermi surface of YbCdSn in 3D-Brillouin Zone without SOC.}
	\label{fig5}
\end{figure}

The bulk band structure of YbCdSn with and without spin-orbit coupling (SOC) calculated for the Yb$^{2+}$ ground state, as established by the SIC-LSDA self-consistent calculations, is shown in Fig.\ref{fig4}. The valence (red) and conduction (blue) bands cross each other along $\Gamma-M$ direction at 0.03 eV and $\Gamma-K$ direction at 0.01 eV as shown  in Fig.\ref{fig4}(a). These bands are constructed by $Yb-d$, $Cd-s$ and $Sn-p$ states and present an inverted band structure at $\Gamma$-M and $\Gamma$-K directions which is supported by character bands of $Cd-s$ and $Sn-p$ orbitals as shown in Fig.\ref{fig4}(c,d). The presence of more than one Dirac points around the same energy level at different high symmetry points indicates the nodal line, and hence the two nontrivial bands, having a constant energy, form a nodal loop above the Fermi level ($E_F$). The (001) projected surface state spectrum of YbCdSn with (Cd,Sn) termination is shown in Fig.\ref{fig5}(a). From this, we can clearly see a nodal line (NL) and surface states (SS) apart from bulk states. The bulk bands touch at Dirac points and surface states merge into bulk states at Dirac points at the same time \cite{wang2017new} which can be seen from Fig.\ref{fig5}(a). The same is confirmed from the Fermi arc calculation in $k_x$-$k_y$ plane which also shows the nodal line pointed by black arrow around $\Gamma$-point as shown in Fig.\ref{fig5}(b) and the other rings may be due to the presence of surface states far from $E_F$ as shown in Fig.\ref{fig5}(a). The three dimensional Fermi surface of YbCdSn without SOC is shown in Fig.\ref{fig5}(c), where one can observe a hexagonal shaped nodal line around $\Gamma$ in $k_x$-$k_y$ plane connecting $M$ and $K$ points. Here the green spherical pocket at the zone centre corresponds to 1 and 2 bands that crossed E$_F$ and the red pockets correspond to band number 3 as shown in Fig.\ref{fig4}(a). With the inclusion of SOC as shown in Fig.\ref{fig4}(b), the degeneracy lifted bands are seen in solid and dashed red and blue lines and the band hybridization opens a gap of 28 $meV $along $\Gamma-K$ direction and tiny gap of 7 $meV $along $\Gamma-M$ direction and the band hybridization is due to $Cd-s$ and $Sn-p$ characters which are shown in Fig.\ref{fig4}(e,f). Now the system becomes topologically nontrivial with the inclusion of SOC, which is confirmed by the Z2 topological indices ($\upsilon_0$; $\upsilon_1$$\upsilon_2$$\upsilon_3$) = (1; 000). The value $\upsilon_0$=1 clearly tells us the nontrivial topology of  the system with the inclusion of SOC.

\section{Discussion and conclusion}

We next turn to discuss the novel transport phenomena and the mass enhancement of the charge carries in the correlated state of YbCdSn. Nearly equal electron and hole densities suggest that electron-hole compensation is responsible for large MR in this compound \cite{CaCdGe_PRB_2017}. The large $\gamma$-value indicates an enhancement of effective mass which is further supported by electronic structure calculation. The calculated effective mass from the electronic structure is around 25.5$m_e$ along $A-L$ and $K-H$ directions. Mass enhancement was also observed in the intermediate valence compound SmB$_6$, where band inversion between opposite parity $4f$ and $5d$ states makes it topologically nontrivial and correlations induced bulk insulating gap formation leads to a topological Kondo insulator state \cite{SmB6_PRL_2014,SmB6_PRL_2013}. In the Weyl semimetal YbPtBi, Kondo coupling strengthens at low temperatures and modifies the dispersion of the nontrivial bands in the vicinity of the topologically protected Weyl points which enhances the effective mass whereas Weyl fermion state dominates at high temperature \cite{YbPtBi_NatureCom_2018}. We have also estimated the Wilson ratio in order to get some idea about the strength of electron correlations. The Wilson ratio is expressed as $R_W= (\pi^2 K_B^2 \chi_0)/(\gamma \mu_{eff})$ \cite{WilsonRatio_1975,YbNiAl4_JAP_2018}. The estimated value of $R_W$ is 0.68 for YbCdSn, where $\chi_0=1.8\times 10^{-3}$ emu/mol, $\mu_{eff}=4.54 \mu_B$ and $\gamma$= 28 mJ mol$^{-1}$ K$^{-2}$. This value of $R_W$ falls in the range of valence fluctuating compounds and indicates significant electron correlations \cite{WilsonRatio_2003,YbNiAl4_JAP_2018}.

In conclusion, our experimental studies reveal that YbCdSn hosts an intermediate valence state. The specific heat indicates an enhancement of effective mass in this state. More intriguing thing is that the correlated phase of this compound shows unusual magnetotransport properties such as large nonsaturating MR of $\sim 2.5 \times 10^2$ \% at 4 K and 12 T, low carrier density, field induced metal-semiconductor-like crossover and a plateau in resistivity at low temperatures. Moreover, a cusp-like feature in magnetoconductivity at low field indicates the presence of WAL effect in this system. On the theory side, our first-principles SIC-LSDA calculations predict a divalent state for Yb ion and the corresponding density of states appears to support the semimetalic behaviour. With applying a considerable pressure, achieved by decreasing volume, a transition from Yb$^{2+}$ to Yb$^{3+}$ was seen to occur at a very small volume, seemingly difficult to realize in practice but not excluding valence fluctuations between the two configurations. As for the band structure calculations, they predict a topological nodal-line semimetalic state of YbCdSn for the ground state divalent Yb ion. To confirm this prediction, and describe valence fluctuations, further experiments such as ARPES and quantum oscillations measurements, as well as calculations accounting for dynamical correlations, would be required.

\section{Acknowledgements}
Research support from IIT Kanpur, IIT Hyderabad and  SERB India (Grant No. CRG/2018/000220) are gratefully acknowledged. This work made use of computational support by CoSeC, the Computational Science Centre for Research Communities, through CCP9.
 
\section{Appendix: SIC-LSD formalism}
The basis of the SIC-LSD formalism is an orbital dependent DFT energy functional,
obtained from the LSDA total energy functional by subtracting a spurious self-interaction of each occupied electron state, thus making it self-interaction free. The self-interaction error is inherent in LSDA due to local approximation applied to the exchange-correlation energy functional and to a large extent is responsible for the failure of LSDA (and even GGA) to describe strong electron correlations. The self-interaction correction (SIC) constitutes a negative energy contribution gained by an electron upon localization which competes with the band formation energy gained by the electron if allowed to delocalize and hybridize with the available conduction electron states. Hence, in this formulation, one distinguishes between localized and itinerant states but treats them on equal footing through expanding them in the same set of basis functions in an {\it ab initio} manner. The decision whether a state is treated as localized or extended is based on a delicate energy balance between band formation and localization. One has to explore a variety of possible configurations consisting of different distributions of localized and itinerant states and minimize the SIC-LSDA total energy functional with respect to all those configurations which leads to a number of local minima of the same functional, meaning that their total energies may be compared. The configuration with the lowest total energy defines the ground state energy and the ensuing valence. If no localized states are assumed, then the SIC-LSDA energy functional coincides with the LSDA energy functional, namely the LSDA functional is also a local minimum of the SIC-LSDA energy functional. The SIC-LSDA formalism still considers the electronic structure of the solid to be built from individual one-electron states, but offers an alternative description to the Bloch picture, namely in terms of periodic arrays of localized atom-centred states, i.e., the Heitler-London picture in terms of exponentially decaying Wannier functions. Nevertheless, there always exist states that will never benefit from SIC, retaining their itinerant character of the Bloch form and moving in the LSDA effective potential. This is the case for the non-$f$ conduction electron states in the lanthanide systems. In the SIC-LSD method, the eigenvalue problem is solved in the space of Bloch states, but a transformation to the Wannier representation is made  at every step of the self-consistency process to calculate the localized orbitals and the corresponding charges to evaluate the SIC potentials of the states that are truly localized. These repeated transformations between Bloch and Wannier representations constitute the major difference between LSDA and SIC-LSDA. Since the main effect of SIC is to reduce hybridization of localized electron states with the valence band, the technical difficulties of minimizing the SIC-LSDA functional in solids are often circumvented by introducing an empirical effective Coulomb interaction parameter U for the orbitals that are meant to be localized, which leads to the DFT+U approach used here for the band structure calculations of YbCdSn. Of course, this way the major advantage of the SIC-LSDA formalism, namely providing a dual picture of coexisting localized and band-like $f$-electrons and allowing one to determine valencies of the constituent elements in the solids, is lost.

\bibliography{ReferenceAll}
\end{document}